\documentstyle[12pt]{article}

\catcode`\@=11
\long\def\@makefntext#1{
\protect\noindent \hbox to 3.2pt {\hskip-.9pt
$^{{\ninerm\@thefnmark}}$\hfil}#1\hfill}		

\def\@makefnmark{\hbox to 0pt{$^{\@thefnmark}$\hss}}  
	
\def\ps@myheadings{\let\@mkboth\@gobbletwo
\def\@oddhead{\hbox{}
\rightmark\hfil\ninerm\thepage}
\def\@oddfoot{}\def\@evenhead{\ninerm\thepage\hfil
\leftmark\hbox{}}\def\@evenfoot{}
\def\sectionmark##1{}\def\subsectionmark##1{}}

\newcounter{sectionc}\newcounter{subsectionc}\newcounter{subsubsectionc}
\renewcommand{\section}[1] {\vspace*{0.6cm}\addtocounter{sectionc}{1}
\setcounter{subsectionc}{0}\setcounter{subsubsectionc}{0}\noindent
	{\normalsize\bf\thesectionc. #1}\par\vspace*{0.4cm}}
\renewcommand{\subsection}[1] {\vspace*{0.6cm}\addtocounter{subsectionc}{1}
	\setcounter{subsubsectionc}{0}\noindent
	{\normalsize\it\thesectionc.\thesubsectionc. #1}\par\vspace*{0.4cm}}
\renewcommand{\subsubsection}[1]
{\vspace*{0.6cm}\addtocounter{subsubsectionc}{1}
	\noindent {\normalsize\rm\thesectionc.\thesubsectionc.\thesubsubsectionc.
	#1}\par\vspace*{0.4cm}}

\newcounter{appendixc}
\newcounter{subappendixc}[appendixc]
\newcounter{subsubappendixc}[subappendixc]

\renewcommand{\appendix}[1] {\vspace*{0.6cm}
        \refstepcounter{appendixc}
        \setcounter{figure}{0}
        \setcounter{table}{0}
        \setcounter{equation}{0}
        \renewcommand{\thefigure}{\Alph{appendixc}.\arabic{figure}}
        \renewcommand{\thetable}{\Alph{appendixc}.\arabic{table}}
        \renewcommand{\theappendixc}{\Alph{appendixc}}
        \renewcommand{\theequation}{\Alph{appendixc}.\arabic{equation}}
        \noindent{\bf Appendix \theappendixc #1}\par\vspace*{0.4cm}}

\def\abstracts#1{{
	\centering{\begin{minipage}{12.2truecm}\footnotesize\baselineskip=12pt\noindent
	\centerline{\footnotesize ABSTRACT}\vspace*{0.3cm}
	\parindent=0pt #1
	\end{minipage}}\par}}

\renewenvironment{thebibliography}[1]
	{\begin{list}{\arabic{enumi}.}
	{\usecounter{enumi}\setlength{\parsep}{0pt}
\setlength{\leftmargin 1.25cm}{\rightmargin 0pt}
	 \setlength{\itemsep}{0pt} \settowidth
	{\labelwidth}{#1.}\sloppy}}{\end{list}}

\newcounter{itemlistc}
\newcounter{romanlistc}
\newcounter{alphlistc}
\newcounter{arabiclistc}

\newcommand{\fcaption}[1]{
        \refstepcounter{figure}
        \setbox\@tempboxa = \hbox{\footnotesize Fig.~\thefigure. #1}
        \ifdim \wd\@tempboxa > 6in
           {\begin{center}
        \parbox{6in}{\footnotesize\baselineskip=12pt Fig.~\thefigure. #1}
            \end{center}}
        \else
             {\begin{center}
             {\footnotesize Fig.~\thefigure. #1}
              \end{center}}
        \fi}

\newcommand{\tcaption}[1]{
        \refstepcounter{table}
        \setbox\@tempboxa = \hbox{\footnotesize Table~\thetable. #1}
        \ifdim \wd\@tempboxa > 6in
           {\begin{center}
        \parbox{6in}{\footnotesize\baselineskip=12pt Table~\thetable. #1}
            \end{center}}
        \else
             {\begin{center}
             {\footnotesize Table~\thetable. #1}
              \end{center}}
        \fi}

\def\@citex[#1]#2{\if@filesw\immediate\write\@auxout
	{\string\citation{#2}}\fi
\def\@citea{}\@cite{\@for\@citeb:=#2\do
	{\@citea\def\@citea{,}\@ifundefined
	{b@\@citeb}{{\bf ?}\@warning
	{Citation `\@citeb' on page \thepage \space undefined}}
	{\csname b@\@citeb\endcsname}}}{#1}}

\newif\if@cghi
\def\cite{\@cghitrue\@ifnextchar [{\@tempswatrue
	\@citex}{\@tempswafalse\@citex[]}}
\def\citelow{\@cghifalse\@ifnextchar [{\@tempswatrue
	\@citex}{\@tempswafalse\@citex[]}}
\def\@cite#1#2{{$\null^{#1}$\if@tempswa\typeout
	{IJCGA warning: optional citation argument
	ignored: `#2'} \fi}}

 1
 1
 1

\font\ninerm=cmr9

\begin{document}

\centerline{\normalsize\bf THE PERTURBATIVE APPROACH TO THE CRITICAL}
\baselineskip=16pt
\centerline{\normalsize\bf BEHAVIOUR OF TWO-MATRIX MODELS IN THE}
\baselineskip=16pt
\centerline{\normalsize\bf LIMIT $N \rightarrow \infty$}

\baselineskip=16pt

\vspace*{0.6cm}
\centerline{\footnotesize S.BALASKA\footnote{On leave of absence from Laboratoire de Physique Th\'eorique, Universit\'e d'Oran, 31100 Es-S\'enia, Algeria, E-mail: balaska@physik.uni-kl.de}  , J. MAEDER\footnote{E-mail: maeder@physik.uni-kl.de} ,  W. R\"UH
L\footnote{E-mail: ruehl@physik.uni-kl.de}}
\baselineskip=13pt
\centerline{\footnotesize\it Department of Physics, University of Kaiserslautern, P.O.Box 3049}
\centerline{\footnotesize\it 67653 Kaiserslautern, Germany}

\vspace*{0.8cm}

\vspace*{3cm}
\abstracts{\normalsize{We construct representations of the Heisenberg algebra by pushing the perturbation expansion to high orders. If the multiplication operators $B_{1,2}$ tend to differential operators of order $l_{2,1}$, respectively, the singularity 
is characterized by $(l_{1},l_{2})$. Let $l_{1} \geq l_{2}$. Then the two cases A : ``$l_{2}$ does not divide $l_{1}$'' and B : ``$l_{2}$ divides $l_{1}$'' need a different treatment. The universality classes are labelled $[p,q]$ where $[p,q]$=[$l_{1}$,$l
_{2}$] in case A and $[p,q]$=[$l_{1}+1$,$l_{2}$] in case B.} }   
\vspace*{1.5cm} \newpage

%
%
\message{reelletc.tex (Version 1.0): Befehle zur Darstellung |R  |N, Aufruf z.B. \string\bbbr}
%
%
%
%
%
\font \smallescriptscriptfont = cmr5
\font \smallescriptfont       = cmr5 at 7pt
\font \smalletextfont         = cmr5 at 10pt
\font \tensans                = cmss10
\font \fivesans               = cmss10 at 5pt
\font \sixsans                = cmss10 at 6pt
\font \sevensans              = cmss10 at 7pt
\font \ninesans               = cmss10 at 9pt
\newfam\sansfam
\textfont\sansfam=\tensans\scriptfont\sansfam=\sevensans
\scriptscriptfont\sansfam=\fivesans
\def\sans{\fam\sansfam\tensans}
\def\bbbr{{\rm I\!R}} 
\def\bbbn{{\rm I\!N}} 
\def\bbbE{{\rm I\!E}} 
\def\bbbm{{\rm I\!M}}
\def\bbbh{{\rm I\!H}}
\def\bbbk{{\rm I\!K}}
\def\bbbd{{\rm I\!D}}
\def\bbbp{{\rm I\!P}}
\def\bbbone{{\mathchoice {\rm 1\mskip-4mu l} {\rm 1\mskip-4mu l}
{\rm 1\mskip-4.5mu l} {\rm 1\mskip-5mu l}}}
\def\bbbc{{\mathchoice {\setbox0=\hbox{$\displaystyle\rm C$}\hbox{\hbox
to0pt{\kern0.4\wd0\vrule height0.9\ht0\hss}\box0}}
{\setbox0=\hbox{$\textstyle\rm C$}\hbox{\hbox
to0pt{\kern0.4\wd0\vrule height0.9\ht0\hss}\box0}}
{\setbox0=\hbox{$\scriptstyle\rm C$}\hbox{\hbox
to0pt{\kern0.4\wd0\vrule height0.9\ht0\hss}\box0}}
{\setbox0=\hbox{$\scriptscriptstyle\rm C$}\hbox{\hbox
to0pt{\kern0.4\wd0\vrule height0.9\ht0\hss}\box0}}}}

\def\bbbe{{\mathchoice {\setbox0=\hbox{\smalletextfont e}\hbox{\raise
0.1\ht0\hbox to0pt{\kern0.4\wd0\vrule width0.3pt height0.7\ht0\hss}\box0}}
{\setbox0=\hbox{\smalletextfont e}\hbox{\raise
0.1\ht0\hbox to0pt{\kern0.4\wd0\vrule width0.3pt height0.7\ht0\hss}\box0}}
{\setbox0=\hbox{\smallescriptfont e}\hbox{\raise
0.1\ht0\hbox to0pt{\kern0.5\wd0\vrule width0.2pt height0.7\ht0\hss}\box0}}
{\setbox0=\hbox{\smallescriptscriptfont e}\hbox{\raise
0.1\ht0\hbox to0pt{\kern0.4\wd0\vrule width0.2pt height0.7\ht0\hss}\box0}}}}

\def\bbbq{{\mathchoice {\setbox0=\hbox{$\displaystyle\rm Q$}\hbox{\raise
0.15\ht0\hbox to0pt{\kern0.4\wd0\vrule height0.8\ht0\hss}\box0}}
{\setbox0=\hbox{$\textstyle\rm Q$}\hbox{\raise
0.15\ht0\hbox to0pt{\kern0.4\wd0\vrule height0.8\ht0\hss}\box0}}
{\setbox0=\hbox{$\scriptstyle\rm Q$}\hbox{\raise
0.15\ht0\hbox to0pt{\kern0.4\wd0\vrule height0.7\ht0\hss}\box0}}
{\setbox0=\hbox{$\scriptscriptstyle\rm Q$}\hbox{\raise
0.15\ht0\hbox to0pt{\kern0.4\wd0\vrule height0.7\ht0\hss}\box0}}}}

\def\bbbt{{\mathchoice {\setbox0=\hbox{$\displaystyle\rm
T$}\hbox{\hbox to0pt{\kern0.3\wd0\vrule height0.9\ht0\hss}\box0}}
{\setbox0=\hbox{$\textstyle\rm T$}\hbox{\hbox
to0pt{\kern0.3\wd0\vrule height0.9\ht0\hss}\box0}}
{\setbox0=\hbox{$\scriptstyle\rm T$}\hbox{\hbox
to0pt{\kern0.3\wd0\vrule height0.9\ht0\hss}\box0}}
{\setbox0=\hbox{$\scriptscriptstyle\rm T$}\hbox{\hbox
to0pt{\kern0.3\wd0\vrule height0.9\ht0\hss}\box0}}}}

\def\bbbs{{\mathchoice
{\setbox0=\hbox{$\displaystyle     \rm S$}\hbox{\raise0.5\ht0\hbox
to0pt{\kern0.35\wd0\vrule height0.45\ht0\hss}\hbox
to0pt{\kern0.55\wd0\vrule height0.5\ht0\hss}\box0}}
{\setbox0=\hbox{$\textstyle        \rm S$}\hbox{\raise0.5\ht0\hbox
to0pt{\kern0.35\wd0\vrule height0.45\ht0\hss}\hbox
to0pt{\kern0.55\wd0\vrule height0.5\ht0\hss}\box0}}
{\setbox0=\hbox{$\scriptstyle      \rm S$}\hbox{\raise0.5\ht0\hbox
to0pt{\kern0.35\wd0\vrule height0.45\ht0\hss}\raise0.05\ht0\hbox
to0pt{\kern0.5\wd0\vrule height0.45\ht0\hss}\box0}}
{\setbox0=\hbox{$\scriptscriptstyle\rm S$}\hbox{\raise0.5\ht0\hbox
to0pt{\kern0.4\wd0\vrule height0.45\ht0\hss}\raise0.05\ht0\hbox
to0pt{\kern0.55\wd0\vrule height0.45\ht0\hss}\box0}}}}

\def\bbbz{{\mathchoice {\hbox{$\sans\textstyle Z\kern-0.4em Z$}}
{\hbox{$\sans\textstyle Z\kern-0.4em Z$}}
{\hbox{$\sans\scriptstyle Z\kern-0.3em Z$}}
{\hbox{$\sans\scriptscriptstyle Z\kern-0.2em Z$}}}}
%
%

\section{Introduction}

Chains of random matrices describe in the double scaling limit $N \rightarrow \infty$ two-dimensional gravity coupled
 to matter which in turn is represented by rational conformal field theories \cite{1,2,3}. A connection of the one-matrix models with the theory of Korteweg-de Vries equations was first established in \cite{2}. In the series of papers \cite{4}$^-$\cite{6}
 the identification of the matter fields with the critical Ising model, Lee-Yang edge singularity,  and the tricritical Ising model was achieved by comparison of the critical exponents (field dimensions). A systematic and constructive approach to arbitrar
y matrix models \cite{7}$^-$\cite{9} was made possible by shifting the interest to representations of a Heisenberg algebra in terms of Gelfand-Dikii quasi-differential operators \cite{10}. This is a special aspect of Korteweg-de Vries equation theory. All
 this progress in matrix model research was obtained by application of the technique of orthogonal polynomials which was developed a long time ago \cite{11,12}. More details of the history of matrix models can be found in review articles (e.g. \cite{13}).

After the completion of the work \cite{7}$^-$\cite{9} interest concentrated on the verification of the universality classes [p,q] where p and q are coprime, in different realizations. From the one-matrix models the sequence of classes [$2l+1$,2], $l \geq 
1$ was obtained and only these. Two-matrix and some three-matrix models were studied in \cite{6},in appendix C of \cite{7}, \cite{8,9} and \cite{14}$^-$\cite{16}. But the universality classes explicitly constructed were not many : [4,3], [5,4], and [8,3].
 The claim that all [p,q] classes appear in two-matrix models was put forward (see the abstract of \cite{9}). This assertion is easily verified by the construction proposed by us.

Our perturbative approach is not based on an investigation of the Heisenberg algebra. On the contrary we construct the representations of this algebra by a perturbative solution of the Schwinger-Dyson equations (synonymous: recursion relations). Generally
 high orders of perturbation theory are needed and a systematic treatment of this expansion is therfore crucial.

The Schwinger-Dyson equations (section 2) involve two multiplication operators $B_{1}$ and $B_{2}$ and two differentiation operators $A_{1}$ and $A_{2}$. We postulate that $B_{1}$ and $B_{2}$ tend to differential operators in the scaling domain
\begin{equation}
B_{1} \rightarrow a^{-l_{2}\gamma}\sum_{n=0}^{\infty} a^{-n\gamma} Q^{(n)}  ,\quad Q^{(0)} = Q
\label{1.1}
\end{equation}
\begin{equation}
B_{2} \rightarrow a^{-l_{1}\gamma}\sum_{n=0}^{\infty} a^{-n\gamma} P^{(n)}  , \quad P^{(0)} = P
\label{1.2}
\end{equation}
If $L_{1}(L_{2})$ is the degree of the potential $V_{1}(V_{2})$ in the two-matrix model action, then necessarily
\begin{equation}
1 \leq l_{i} \leq L_{i}
\label{1.3}
\end{equation}
All $\{Q^{(n)}\}$, $\{P^{(n)}\}$ are differential operators of order
\begin{equation}
\mbox{ord}\, Q^{(n)} = \mbox{ord}\, Q + n = l_{2} + n 
\label{1.4}
\end{equation}
\begin{equation}
\mbox{ord}\, P^{(n)} = \mbox{ord}\, P + n = l_{1} + n
\label{1.5}
\end{equation}
We shall always assume $l_{1} \geq l_{2}$. $\gamma$ is the ``string susceptibility exponent'' and is negative,
\begin{equation}
\frac{1}{N} = a^{2-\gamma}
\label{1.6}
\end{equation}
and $N$ tends to infinity.

If the ansatz (\ref{1.1}), (\ref{1.2}) is satisfied we speak of a ``singularity of type ($l_{1}$,$l_{2}$)'' in the action of the model. We can choose this singularity ``maximal'' by setting
\begin{equation}
l_{1} = L_{1}  ,  l_{2} = L_{2}
\label{1.7}
\end{equation}
We are convinced that all types of singularities that exist can be found if we let $L_{1}$, $L_{2}$ run over all naturals.

After solution of the Schwinger-Dyson equations the commutator
\begin{equation}
[B_{1},A_{1}] = [B_{2},B_{1}] = [A_{2},B_{2}]
\label{1.8}
\end{equation}
is diagonal. Setting it equal the unit operator gives one nonlinear differential equation. The perturbative expansion starts from a linear system of equations for the deviations of $B_{1}$ , $B_{2}$ from their limiting difference operator. This linear sys
tem has corank $l_{2}-2$. It uses two functions $u_{1}$ , $v_{-1}$ as basis functions. Use of biorthogonal systems of polynomials introduces a gauge degree of freedom that affects the whole perturbation theory. The ``susceptibility function'' $u$ is gauge
 invariant
\begin{equation}
u = u_{1} + v_{-1} + a^{-2\gamma}u_{1}v_{-1}
\label{1.9}
\end{equation}
(this form for $u$ is valid in our preferred normalization, see section 3).
The nonvanishing corank $l_{2}-2$ permits the introduction of $l_{2}-2$ new gauge invariant basis functions $\{w_{i}\}_{i=1}^{l_{2}-2}$. Higher order perturbation theory gives one nonlinear differential equation for each one, and these equations are ident
ical to those obtained from the commutator (8) if we postulate that it be diagonal.

After we study several examples in detail we can summarize our results as follows. There are two series of singularity types :

(A) $l_{2}$ does not divide $l_{1}$ (remember $l_{1} \geq l_{2}$).
Then only the leading orders in (\ref{1.1}),(\ref{1.2}) are relevant and in the sense of quasi-differential operators \cite{10}
\begin{equation}
(-)^{l_1} P = Q^{l_{1}/l_{2}}_{+}
\label{1.10}
\end{equation}
The susceptibility exponent is
\begin{equation}
\gamma = \frac{-2}{p+q-1}\
\label{1.11}
\end{equation}
where the universality class is
\begin{equation}
[p,q] = [l_{1},l_{2}]
\label{1.12}
\end{equation}

(B) $l_{2}$ divides $l_{1}$, but $l_{2} \neq 2$. Then
\begin{equation}
(-)^{l_{1}} \, P = Q^{l_{1}/l_{2}}
\label{1.13}
\end{equation}
and 
\begin{equation}
[B_{2},B_{1}] \rightarrow a^{-(l_{1}+l_{2}+1)\gamma}\; [\tilde{P} , Q] + \mbox{ higher order terms}
\label{1.14}
\end{equation}
where
\begin{equation}
\tilde{P} = (-)^{l_{1}}\,P^{(1)} - \sum_{1 \leq m \leq M } Q^{M-m} Q^{(1)} Q^{(m-1)}\quad (M = \mbox{$\frac{l_{1}}{l_{2}}$})
\label{1.15}
\end{equation}
We show that
\begin{equation}
\tilde{P} = c(l_{1},l_{2}) \, Q^{\frac{l_{1}+1}{l_{2}}}_{+}
\label{1.16}
\end{equation}
with
\begin{equation}
c(l_{1},l_{2}) = \frac{(l_{1}l_{2}-l_{1}-l_{2})}{l_{2}}
\label{1.17}
\end{equation}
and correspondingly that
\begin{equation}
\gamma = \frac{-2}{p+q-1}
\label{1.18}
\end{equation}
with
\begin{equation}
[p,q] = [l_{1}+1 , l_{2}]
\label{1.19}
\end{equation}

Some comments are in order in this context. First there is no universality class for singularities $(2m , 2)$. We will later in (section 5) give an argument why this is so. Next a universality class with
\begin{equation}
p = nq + 1 \; , \; n \in \bbbn
\label{1.20}
\end{equation}
can be realized by a singularity either in (A) or in (B). Finally universality classes where $p$ and $q$ have a common divisor but this divisor is not $q$ itself, are contained in (A). The symmetric case $l_{1}=l_{2}$ belongs to series (B).

We study the double scaling limit with $N \rightarrow \infty$ and only one coupling tuned to a critical value. Nevertheless the complete set of nonlinear differential equations for $u$ and $\{w_{i}\} $ is derived. Only the ``integration constants'' are ze
ro and $x$ is the single variable in the equations that is kept.

In order to make this work self-contained we introduce matrix models and discuss their analysis by the orthogonal polynomial method in section 2. Two elementary and easily proved propositions which are not new but play a crucial role in the perturbative a
pproach are given the form of ``Lemmas''. In section 3  we develop the perturbative analysis of the Dyson-Schwinger equations systematically, the decisive observation being that the corank of the linear approximation is $l_{2} - 2$. This method is complet
e in the sense that any singularity $(l_{1},l_{2})$ can be treated with it. In the subsequent two sections we study examples from the series (A) : $(4,3)$ , $(5,4)$ and $(6,4)$; and from the series (B) : $(3,3)$ and $(6,3)$. The two examples $(4,3)$ and $
(5,4)$ (where $l_{1}$ and $l_{2}$ are coprime) serve only the purpose to prove that our method reproduces exactly the known results. Moreover we show in section 5 that singularities $(2m,2)$ do not fulfill the commutator equation. Expectation values, in p
articular their role in tests of gauge invariance, are studied in section 6.

\section{The orthogonal polynomial approach to matrix models}

Matrix models based on $N \times N$ hermitean or unitary or other matrices
as dynamical variables have a history of about twenty years. It has turned 
out that a class of ``solvable'' matrix models exist which are built on a 
finite number $r$ of such matrices
\begin{equation} \label{2.1}
\{ M^{(\alpha)} \}
\end{equation}
which are bilinearly coupled to a \underline{chain}. In the case of
hermitean matrix models (to which we shall restrict our attention) this means
an action
\begin{equation} \label{2.2}
S(M^{(1)}, M^{(2)}, \ldots, M^{(r)}) = \mbox{Tr} \left\{ \sum_{\alpha=1}^{r} 
V_{\alpha}(M^{(\alpha)}) - \sum_{\alpha=1}^{r-1} c_{\alpha} M^{(\alpha)}
M^{(\alpha+1)} \right\}
\end{equation}
where each $V_{\alpha}$ is a polynomial of degree $L_{\alpha}$ (only 
$L_{\alpha} \geq 3$ is of interest)
\begin{equation} \label{2.3}
V_{\alpha}(x) = \sum_{k=1}^{L_{\alpha}} \frac{g_{k}^{(\alpha)}}{k} x^{k}
\end{equation}
The action is stable if and only if
\begin{equation} \label{2.4}
\left\{ {L_{\alpha} \mbox{ is even, all } \alpha \atop 
g_{L_{\alpha}}^{(\alpha)} \mbox{ is positive, all } \alpha} \right.
\end{equation}
In the stable case the partition function of the model is
\begin{eqnarray}
\label{2.5} Z &=& \int \prod_{\alpha} dM^{(\alpha)} e^{-S} \\
\label{2.6} dM^{(\alpha)} &=& \prod_{i \leq j} d(\mbox{Re}M_{ij}^{(\alpha)})
\prod_{k < l} d(\mbox{Im}M_{kl}^{(\alpha)})
\end{eqnarray}

If the action is symmetric under
\begin{equation} \label{2.7}
M^{(\alpha)} \leftrightarrow M^{(r+1-\alpha)}
\end{equation}
we call the model ``$\bbbz_{2}$-symmetric'', otherwise ``asymmetric''. A model may
have a $\bbbz_{2}$-symmetric critical point which can be approached in an
asymmetric fashion such as an Ising model with external field $H$ at the
critical point.

When we claim that such matrix models (\ref{1.2}) are ``solvable'', we mean 
solvability in the scaling domain of a critical point which implies the 
existence of a full asymptotic series
\begin{equation} \label{2.8}
\langle \Omega \rangle \sim \sum_{m=0}^{\infty} a^{-m \gamma} \omega_{m}(z)
\end{equation}
with $a^{- \gamma} \rightarrow 0$ in the scaling domain, $z$ a scale invariant
variable and computability of all $\omega_{m}$ by analytic methods, for any 
``observable'' $\Omega$. These observables include thermodynamic objects such as
$\log Z$ or correlation functions of the ``Mehta-accessible'' class (see section
6). Our aim is to develop an algorithm so far, that for a given action all
singularities can be extracted, the universality class can be described for 
each case and non-leading orders in the scaling behaviour can be calculated
to any precision desired.

The full richness of structure develops only in asymmetric two-matrix and 
many-matrix models, which have not been studied carefully before. The symmetric
models can often be obtained thereafter by a reduction of degrees of freedom.
In any case we apply the method of orthogonal polynomials (by Bessis and Mehta,
\cite{11,12}). By a theorem of Mehta the action (\ref{2.2}) can be 
expressed in terms of eigenvalues
\begin{equation} \label{2.9}
\{ \lambda_{i}^{(\alpha)} \}_{1}^{N} \: \mbox{ of } \: M^{(\alpha)} 
\end{equation} 
\begin{equation} \label{2.10}
S(\{ \lambda_{i}^{(\alpha)} \}) = \sum_{i=1}^{N} \left[ \sum_{\alpha=1}^{r} 
V_{\alpha} (\lambda_{i}^{(\alpha)}) - \sum_{\alpha=1}^{r-1} c_{\alpha} 
\lambda_{i}^{(\alpha)} \lambda_{i}^{(\alpha+1)} \right]
\end{equation} 
The partition function goes into 
\begin{equation} \label{2.11}
Z = C_{r}(N) \int \prod_{\alpha=1}^{r} d\mu(\Lambda^{(\alpha)}) \Delta(
\lambda^{(1)}) \Delta(\lambda^{(r)}) \exp[-S( \{ \lambda_{i}^{(\alpha)} \})]
\end{equation}
where the measure is
\begin{equation} \label{2.12}
d\mu(\Lambda^{(\alpha)}) = \prod_{i=1}^{N} d\lambda_{i}^{(\alpha)},
\end{equation}
and
\begin{equation} \label{2.13}
\Delta(\lambda) = \prod_{i<j} (\lambda_{i}-\lambda_{j})
\end{equation}
is the Vandermonde determinant. For the solvability it is crucial that both
$\Delta(\lambda)$ appear in the numerator.

The biorthogonal polynomials

\begin{equation} \label{2.14}
\{ \Pi_{m}(\lambda), {\tilde{\Pi}}_{n}(\mu) \}_{m,n=0}^{\infty}
\end{equation}
are defined by
\begin{equation} \label{2.15}
\deg \Pi_{m} = \deg {\tilde{\Pi}}_{m} = m
\end{equation}
and
\begin{equation} \label{2.16}
\int \prod_{\alpha=1}^{r} d\lambda^{(\alpha)} \Pi_{m}(\lambda^{(1)}) 
{\tilde{\Pi}}_{n}(\lambda^{(r)}) \exp \{ -\sum_{\alpha=1}^{r} 
V_{\alpha} (\lambda^{(\alpha)}) + \sum_{\alpha=1}^{r-1} c_{\alpha} 
\lambda^{(\alpha)} \lambda^{(\alpha+1)} \} = \delta_{mn}
\end{equation}
through the Schmidt orthogonalization procedure. But this procedure determines
only $2m+1$ of the $2m+2$ coefficients in $\Pi_{m}, \; {\tilde{\Pi}}_{m}$. In
fact we have the \\
\underline{First Lemma:} \\
Let
\begin{eqnarray}
\label{2.17} \Pi_{m}(\lambda) &=& s_{m} \lambda^{m} + \mbox{O}(\lambda^{m-1}) \\
\label{2.18} {\tilde{\Pi}}_{m}(\mu) &=& {\tilde{s}}_{m} \mu^{m} + \mbox{O}(\mu^{m-1})
\end{eqnarray}
then the orthogonalization procedure determines only
\begin{equation} \label{2.19}
s_{m} {\tilde{s}}_{m}
\end{equation}
but not $s_{m}$ and ${\tilde{s}}_{m}$ separately. $\Box$

This leads to the following
corollary: We can of course require that
\begin{equation} \label{2.20}
s_{m} = {\tilde{s}}_{m} > 0
\end{equation}
as in the $\bbbz_{2}$-symmetric case, but expectation values of observables
ought to depend only on $s_{m} {\tilde{s}}_{m}$ and not on such arbitrary
definitions. They do, as we shall see, but in a nontrivial fashion (related
with the gauge invariance of Yang-Mills theory observables).

We follow the usual line and introduce differentiation matrices
\begin{eqnarray}
\label{2.21} {\Pi'}_{m} &=& \sum_{n} (A_{1})_{mn} \Pi_{n} \\
\label{2.22} {\tilde{\Pi}'}_{m} &=& \sum_{n} (A_{r})_{nm} {\tilde{\Pi}}_{n}
\end{eqnarray}
and $r$ multiplication matrices $B_{\alpha}$
\begin{eqnarray}
\label{2.23} \lambda \, {\Pi}_{m} &=& \sum_{n} (B_{1})_{mn} \Pi_{n}(\lambda) \\
\label{2.24} \mu \, {\tilde{\Pi}}_{m} &=& \sum_{n} (B_{r})_{nm} 
{\tilde{\Pi}}_{n}(\mu)
\end{eqnarray}
\begin{center} ($B_2, B_3, \ldots, B_{r-1}$ are not given here) \end{center}
so that
\begin{equation} \label{2.25}
[B_{1}, A_{1}] = [A_{r}, B_{r}] = 1
\end{equation}
By arguments typical for the derivation of Dyson-Schwinger equations one can
derive a set of ``equations of motion''
\begin{eqnarray}
A_{1} + c_{1} B_{2} &=& {V'}_{1}(B_{1}) \nonumber \\
 &\vdots& \nonumber \\
c_{\alpha-1} B_{\alpha-1} + c_{\alpha} B_{\alpha+1} &=& {V'}_{\alpha}(B_{\alpha}),
\quad 2 \leq \alpha \leq r-1 \nonumber \\
 &\vdots& \nonumber \\
\label{2.26} A_{r} + c_{r-1} B_{r-1} &=& {V'}_{r}(B_{r}).
\end{eqnarray}
If these equations are satisfied then
\begin{eqnarray}
[B_{1}, A_{1}] &=& c_{1} [B_{2}, B_{1}] \nonumber \\
\cdots &=& c_{\alpha} [B_{\alpha+1}, B_{\alpha}] = \cdots \nonumber \\
\label{2.27} &=& c_{r-1} [B_{r}, B_{r-1}] \:\, = [A_{r}, B_{r}]
\end{eqnarray}
can be shown to hold easily. Moreover the following restrictions on the domain
of $A_{1}, A_{r}, B_{\alpha}$, where these matrices have nonzero entries, can
be shown to be consistent with the Dyson-Schwinger equations:
\begin{eqnarray}
\label{2.28} (A_{1})_{m,n} &=& 0 \; \mbox{ except for } \; 
-\!\prod_{\alpha=1}^{r} (L_{\alpha}-1) \leq n-m \leq -1 \\
\label{2.29} (A_{r})_{m,n} &=& 0 \; \mbox{ except for } \qquad \qquad \qquad
1 \leq n-m \leq \prod_{\alpha=1}^{r} (L_{\alpha}-1) \\
\label{2.30} (B_{\alpha})_{m,n} &=& 0 \; \mbox{ except for } \;
-\!\prod_{\beta>\alpha}^{r} (L_{\beta}-1) \leq n-m \leq
\prod_{\beta<\alpha}^{r} (L_{\beta}-1)
\end{eqnarray}
Our strategy is to solve the Dyson-Schwinger equations (\ref{2.26}) 
perturbatively under the constraints (\ref{2.28})-(\ref{2.30}) and impose the
commutator equations (\ref{2.25}) at the end. This approach is considerably
simplified by the following \\
\underline{Second Lemma:} \\
If (\ref{2.26}) and (\ref{2.28})-(\ref{2.30}) hold, then the commutators
(\ref{2.27}) are diagonal.

To prove this lemma we remind the reader that the chain (\ref{2.27}) of
equations follows from (\ref{2.26}). On the other hand from (\ref{2.28})-
(\ref{2.30}) one can conclude that \\
\indent $[B_{1}, A_{1}]$ is a lower left triangular matrix and \\
\indent $[A_{r}, B_{r}]$ is an upper right triangular matrix. \\
Equality of both commutators proves that they are diagonal.

Then we can express the equation (\ref{2.25}) by
\begin{eqnarray}
(B_{1} A_{1})_{nn} &=& c_{1} (B_{2} B_{1})_{nn} = \cdots \nonumber \\
\label{2.31} &=& n + \mbox{ const. indep. of $n$}
\end{eqnarray}

\section{The perturbative expansion in the double scaling domain}

Those coupling constants that appear as expansion coefficients in the 
potentials $V_{\alpha}$ (\ref{2.2}), (\ref{2.3}) are fixed to critical
values. The corresponding critical action is then multiplied with
\begin{equation} \label{3.1}
\frac{N}{g}
\end{equation}
and tuned as
\begin{equation} \label{3.2}
N \rightarrow \infty, \quad g \rightarrow g_c.
\end{equation}
The matrix labels $n, m$ become continuous variables in this limit
\begin{eqnarray}
\frac{n}{N} &=& \xi \nonumber \\
\label{3.3} 0 &<& \xi \leq 1 \quad \mbox{(for the labels of the eigenvalues).} 
\end{eqnarray}
With the help of the string susceptibility exponent $\gamma \; (0 < -\gamma
< \frac{1}{2})$ one defines
\begin{eqnarray}
\label{3.4} \frac{1}{N} &=& a^{2-\gamma} \\
\label{3.5} \xi &=& \frac{g_c}{g} (1-a^{2} x) \\
\label{3.6} g &=& g_c (1-a^{2} z) \frac{}{} \\
\label{3.7} z &=& (g_c -g) N^{\frac{2}{2-\gamma}} \frac{}{};
\end{eqnarray}
$z$ is the single scale invariant variable, later we will introduce the 
``integration constants'' $h_s$ that play a similar role, and $\gamma$ is as in
(\ref{1.11}), (\ref{1.18}).

For the two-matrix models, be they symmetric or not, we proceed as follows:
We set
\begin{eqnarray}
\label{3.8} (B_{1})_{n,n+k} &=& r_{k}(n) \\
\label{3.9} (B_{2})_{n,n+k} &=& s_{k}(n)
\end{eqnarray}
and define
\begin{eqnarray}
\label{3.10} r_{k}(n-\mbox{$\frac{k}{2}$}) &=& \rho_{k} + a^{-2 \gamma} 
u_{k}(x) \\
\label{3.11} s_{k}(n-\mbox{$\frac{k}{2}$}) &=& \sigma_{k} + a^{-2 \gamma} 
v_{k}(x).
\end{eqnarray}
The translation of the arguments of $r_k, s_k$ symmetrizes the support of the
matrices $B_1, B_2$ around the diagonal and saves labour (and terms) in the
perturbative expansions. For the coefficients $\{ \rho_k, \sigma_k \}$ we make
the ansatz
\begin{eqnarray}
\label{3.12} \sum_{k=-(l_{2}-1)}^{1} \rho_k z^{k} &=&  z \left( 1-
\frac{1}{z} \right)^{l_{2}} \\
\label{3.13} \sum_{k=-1}^{l_{1}-1} \sigma_k z^{k} &=& \frac{1}{z} 
(1-z)^{l_{1}}
\end{eqnarray}
which is in agreement with the support conditions (\ref{2.28})-(\ref{2.30})
if we set $l_1=L_1, \, l_2=L_2$ so that the singularity is maximal. Moreover
we set
\begin{equation} \label{3.14}
c_1 =1
\end{equation}
in (\ref{2.26}). The normalizations (\ref{3.12})-(\ref{3.14}) can be changed;
these renormalizations will be discussed at the end of this section.

Inserting (\ref{3.12})-(\ref{3.14}) into the Dyson-Schwinger equations 
(\ref{2.26}) gives critical coupling constants
\begin{eqnarray}
\label{3.15} g_{k}^{(1)} &\rightarrow& g_{k}^{(l_1, l_2)} \\
\label{3.16} g_{k}^{(2)} &\rightarrow& g_{k}^{(l_2, l_1)}
\end{eqnarray}
which can be presented in the explicit form
\begin{eqnarray}
(-1)^{k} g_{k}^{(l_1, l_2)} &=& {l_1 \choose k} - \sum_{l_1 \geq m > k}
{l_1 \choose m} {l_2 (m-1) \choose m-k} \nonumber \\
&+& \sum_{l_1 \geq m > m' > k} {l_1 \choose m} {l_2 (m-1) \choose m-m'} 
{l_2 (m'-1) \choose m'-k} \nonumber \\
\label{3.17} &\pm& \ldots
\end{eqnarray}
The Dyson-Schwinger equations can then be given as two systems of equations
for the $\{ v_{k}(x) \}$ and $\{ u_{-k}(x) \}$
\begin{eqnarray}
\label{3.18} v_{k}(x) &=& S_{k}([u_{-m}(x)], \{ g_{r}^{(l_1, l_2)} \}), \quad
0 \leq k \leq l_1-1 \\
\label{3.19} u_{-k}(x) &=& R_{-k}([v_{m}(x)], \{ g_{r}^{(l_2, l_1)} \}), \quad
0 \leq k \leq l_2-1.
\end{eqnarray}
We first observe that there exist no equations for $v_{-1}(x)$ and $u_{1}(x)$, 
so that all \\ 
$\{ v_{k}(x), u_{-k}(x) \}$ are expressed in terms of these functions and 
their derivatives (plus the $\{ w_{k}(x) \}$). Second we look at the linear approximation to (\ref{3.18}), (\ref{3.19})
\begin{eqnarray}
\label{3.20} v_{k}(x) - \sum_{m=-1}^{l_2-1} \Omega_{km}^{(l_1, l_2)} u_{-m}
&=& 0 \\
\label{3.21} u_{-m}(x) - \sum_{k=-1}^{l_1-1} \Omega_{mk}^{(l_2, l_1)} v_{k}
&=& 0
\end{eqnarray}
and find that its corank is $l_2-2$ (remember $l_{2} \leq l_{1}$). The general form of the matrix 
$\Omega^{(l_1, l_2)}$ is
\begin{equation} \label{3.22}
\Omega_{km}^{(l_1, l_2)} = (-1)^{k+m} \sum_{n=k+m+2}^{l_1} (n-1)(-1)^{n}
g_{n}^{(l_1, l_2)} {l_2 (n-2) \choose n-(k+m+2)}.
\end{equation}
Obviously $\Omega_{km}^{(l_1, l_2)}$ depends on only $k+m$. Due to the nonvanishing
corank ($l_2 >2$) we need $l_2-2$ additional functions $\{ w_{k} \}$ which in
the perturbative expansion of the Schwinger-Dyson equations appear each at 
order $a^{-k \gamma}$. Explicitly this expansion looks as
\begin{eqnarray}
\left( \begin{array}{c} v_0 \\ v_1 \\ v_2 \\ \vdots \\ v_{l_1-1} \end{array}
\right) &=& S^{(0)} {u_1 \choose v_{-1}} + a^{-\gamma} S^{(1)} w_1 +
a^{-2 \gamma}
S^{(2)} \left( \begin{array}{c} u_1'' \\ v_{-1}'' \\ u_1^2 \\ u_1 v_{-1} \\
v_{-1}^2 \\ w_2 \end{array} \right) \nonumber \\
\label{3.23} &+& a^{-3 \gamma} S^{(3)} \left( \begin{array}{c} w_1'' \\ w_1
u_1 \\ w_1 v_{-1} \end{array} \right) + \mbox{O}(a^{-4 \gamma})
\end{eqnarray}
and correspondingly
\begin{equation} \label{3.24}
\left( \begin{array}{c} u_0 \\ u_{-1} \\ u_{-2} \\ \vdots \\ u_{-l_2+1} 
\end{array} \right) = R^{(0)} {u_1 \choose v_{-1}} + \ldots
\end{equation}
The matrices $S^{(n)}, R^{(n)}$ have rational numbers as entries.

On the other hand the Dyson-Schwinger equations imply that these new basis
functions satisfy (to leading order) nonlinear differential equations
\begin{equation} \label{3.25}
D_{s}( \{ w_k \} ; u_1, v_{-1} ) = 0, \qquad 1 \leq s \leq l_2-2.
\end{equation}
The introduction of the $\{ w_k \} $ is not unambiguous, e.g. one can add to
$w_2$ any linear combination of the functions appearing in the same column in
(\ref{3.23}) as $w_2$ itself or any higher order function such as 
$a^{-\gamma} .w_1''$. The set of all possible choices can be reduced by the
requirement of gauge invariance (section 6) .

Since we have recursion relations for $s_m, {\tilde{s}}_m$ (\ref{2.17}), 
(\ref{2.18})
\begin{eqnarray}
\label{3.26} \frac{s_m}{s_{m+1}} &=& r_{1}(m) \\
\label{3.27} \frac{{\tilde{s}}_m}{{\tilde{s}}_{m+1}} &=& s_{-1}(m+1)
\end{eqnarray}
the First Lemma implies that all observables may depend only on
\begin{equation} \label{3.28}
r_{1}(n-\mbox{$\frac{1}{2}$}) s_{-1}(n+\mbox{$\frac{1}{2}$}) = 1 + 
a^{-2 \gamma} (u_{1}(x)+ v_{-1}(x)) + a^{-4 \gamma} u_{1}(x) v_{-1}(x)
\end{equation}
and other gauge invariant functions. Thus we are led to define as in 
(\ref{1.9})
\begin{equation} \label{3.29}
u(x) = u_{1}(x) + v_{-1}(x) + a^{-2 \gamma} u_{1}(x) v_{-1}(x).
\end{equation}
We emphasize that (\ref{3.28}), (\ref{3.29}) depend on our normalization
(\ref{3.12}), (\ref{3.13})
\begin{equation} \label{3.30}
\rho_1 = \sigma_{-1} = 1
\end{equation}
and must be changed if we renormalize these quantities (see below). Thus
we conclude that also the functions $\{ w_k \}_{1}^{l_2-2}$ must be introduced
in such a way that the free energy, all expectation values, the leading orders
of the differential operators in (\ref{1.8}) and the equations (\ref{3.25})
depend only on $u$ and $\{ w_k \}$.

Now we rescale $N$ such that
\begin{equation} \label{3.31}
a^{-\gamma} \rightarrow \Lambda a^{-\gamma}.
\end{equation}
It is obvious from (\ref{3.10}), (\ref{3.11}) and (\ref{3.23}), (\ref{3.24}),
(\ref{3.29}) that to all functions $u_k, v_k, u, w_k$ and their derivatives
can be ascribed a dimension (or ``degree'', ``grade'') under (\ref{3.31})
\begin{eqnarray}
\label{3.32} \mbox{dim} \{ u_k, v_k, u \} &=& 2 \quad \quad \mbox{(all $k$)} 
\frac{}{} \\
\label{3.33} \mbox{dim} \{ w_k \} &=& k+2 \frac{}{} \\
\label{3.34} \mbox{dim} \left\{ \frac{d}{dx} \right\} &=& 1.
\end{eqnarray}
Let us consider $c_1$ as a free parameter in the Dyson-Schwinger equations
but maintain the normalizations (\ref{3.12}), (\ref{3.13}). Then the critical
coupling constants (\ref{3.17}) are replaced by
\begin{equation} \label{3.35}
g_{k}^{(l_1, l_2)} \rightarrow c_1 g_{k}^{(l_1, l_2)}
\end{equation}
but the perturbative solution of the Dyson-Schwinger equations is otherwise
unchanged. The asymptotic expressions (\ref{1.1}), (\ref{1.2}) for $B_1, B_2$
remain the same and the commutator equation is now
\begin{equation} \label{3.36}
c_1 \left[ \sum_{n} a^{-n \gamma} P^{(n)}, \sum_{m} a^{-m \gamma} Q^{(m)} 
\right] = \left\{ {1 \quad \mbox{ (case A)} \atop a^{-\gamma} \mbox{ (case B)}} 
\right. 
\end{equation}
Using the concept of dimension in the above mentioned sense and the dimensional
homogeneity of the differential equations (\ref{3.25}), (\ref{3.36}), we can
show that $c_1$ can be tuned to one by an appropriate rescaling.

We can do more. Renormalizing (\ref{3.12}), (\ref{3.13}) by replacing the
r.h.s. by
\begin{equation} \label{3.37}
d_1 z \left( 1-\frac{1}{z} \right)^{l_2} \; \mbox{ resp. } \;
d_2 \frac{1}{z} (1-z)^{l_1}
\end{equation}
and keeping $c_1$ as in (\ref{3.14}), we obtain a new gauge invariant function
for which (\ref{1.9}), (\ref{3.29}) is a special case
\begin{equation} \label{3.38}
u(x) = d_2 u_{1}(x) + d_1 v_{-1}(x) + a^{-2 \gamma} u_{1}(x) v_{-1}(x).
\end{equation}
This eliminates one degree of freedom in the renormalization. The remaining
degree of freedom can be eliminated by a rescaling just as $c_1$. Thus we have
shown that no relevant free parameters are left over in the characterization
of the universality classes.

In case (A) the commutator
\begin{eqnarray} \label{3.39}
&[P, Q] = \sum_{s=0}^{l_{2}-2} \{ r_s, \partial^{s} \}, \quad 
\partial=\frac{d}{dx} & \nonumber \frac{}{} \\ & (P=P^{(0)}, Q=Q^{(0)}) &
\end{eqnarray}
does not vanish identically and can therefore be set equal to 1. Now the
second Lemma asserts that
\begin{equation} \label{3.40}
r_s = 0, \quad s \in \{1,2, \ldots, l_2-2 \}
\end{equation}
whereas
\begin{equation} \label{3.41}
2 r_0 = 1
\end{equation}
holds by fiat. The equations (\ref{3.40}) are integrable to
\begin{equation} \label{3.42}
\int r_s \, dx = h_s, \quad s \in \{1,2, \ldots, l_2-2 \}.
\end{equation}
The ``constants of integration'' $h_s$ play the role of external scaling fields.
Equation (\ref{3.41}) can be integrated to
\begin{equation} \label{3.43}
2 \int r_0 \, dx = x.
\end{equation}
The differential equations determine $u$ and all $w_k$ as functions of $x$
and $h_s$. This is the standard approach \cite{7}.

On the other hand we have in all cases studied that
\begin{equation} \label{3.44}
D_s ( w_k, u) = \int r_s \, dx = 0.
\end{equation}
Since we have tuned only one coupling constant g, the integration constants
are automatically zero. Of course we can also introduce effective actions,
e.g.
\begin{equation} \label{3.45}
S_{\mbox{\scriptsize{eff}}} = \mbox{Tr} \left\{ Q^{\frac{l_1+l_2}{l_2}}+
\sum_{s=0}^{l_2-2} t_s Q^{\frac{s+1}{l_2}} \right\}
\end{equation}
for case (A), where
\begin{equation} \label{3.46}
t_0 \sim x, \quad t_s \sim h_s \quad (s \geq 1)
\end{equation}
Thus the known picture (as presented in \cite{7}, say) is completely reproduced
by our perturbative approach.

\section{Examples from the series (A)}

In this and the subsequent section we study the differential operator limits of $B_{1}$ and $B_{2}$. For this purpose we need to know the numbers
\begin{equation}
P_{n} = \sum_{k} k^n \rho_{k}
\label{4.1}
\end{equation}
\begin{equation}
\Sigma_{n} = \sum_{n} k^n \sigma_{k}
\label{4.2}
\end{equation}
and the functions
\begin{equation}
U_{n} = \sum_{k} k^n u_{k}
\label{4.3}
\end{equation}
\begin{equation}
V_{n} = \sum_{k} k^n v_{k}
\label{4.4}
\end{equation}
From (\ref{3.12}),(\ref{3.13}) results
\begin{eqnarray}
P_{n} & = & 0 , \quad n < l_{2} \label{4.5} \\
P_{l_{2}} & = &  l_{2}! \label{4.6} \\
P_{l_{2}+1} & = & (1 - \frac{l_{2}}{2})(l_{2} + 1)! \label{4.7}
\end{eqnarray}
and
\begin{eqnarray}
\Sigma_{n} &=& 0 , \quad n < l_{1} \label{4.8} \\
\Sigma_{l_{1}} &=& (-)^{l_{1}} \, l_{1}! \label{4.9} \\
\Sigma_{l_{1}+1} &=& (-)^{l_{1}+1} (1 - \frac{l_{1}}{2}) (l_{1}+1)! \label{4.10}
\end{eqnarray}
In terms of these quantities the asymptotic expansions of $B_{1}$ and $B_{2}$ are
\begin{equation}
a^{l_{2} \gamma}B_{1} = \sum_{r=0}^{\infty} q_{r}(x) \partial^{r} = \sum^{\infty}_{n=0} a^{-n \gamma} Q^{(n)}
\label{4.11}
\end{equation}
\begin{equation}
a^{l_{1} \gamma}B_{2} = \sum_{s=0}^{\infty} p_{s}(x) \partial^{s} = \sum^{\infty}_{n=0} a^{-n \gamma} P^{(n)}
\label{4.12}
\end{equation}
where
\begin{equation}
q_{r}(x) = \frac{1}{r!} [a^{-(r-l_{2})\gamma}\, P_{r} + \sum_{n=0}^{\infty} a^{-(n+2-l_{2}) \gamma}\; \frac{U^{(n)}_{n+r}(x)}{2^n n!}]
\label{4.13}
\end{equation}
\begin{equation}
p_{s}(x) = \frac{1}{s!} [a^{-(s-l_{1})\gamma}\, \Sigma_{s} + \sum_{n=0}^{\infty} a^{-(n+2-l_{1}) \gamma}\; \frac{V^{(n)}_{n+s}(x)}{2^n n!}]
\label{4.14}
\end{equation}
For the series (A) only $ Q = Q^{(0)}$ and $P = P^{(0)}$ are of interest.

In the case $(l_{1},l_{2}) = (4,3)$ which describes the critical Ising model, we have only one function $w=w_{1}$ which we define by (exact to all orders)
\begin{equation}
u_{0} = u_{1} + v_{-1} + a^{-\gamma}w_{1} 
\label{4.15}
\end{equation}
It turns out that this definition of $w_{1}$ can be chosen in all models (see, however, (\ref{6.17})). We obtain from (\ref{3.23}), (\ref{3.24})
\begin{eqnarray}
U_{0} & = & 3 w \, a^{-\gamma} + O(a^{-2 \gamma}) \label{4.16} \\
U_{1} & = & 3 u + O(a^{- \gamma}) \label{4.17} \\
U_{2} & = & O(1) \label{4.18} \\
V_{0} & = & (\mbox{$\frac{2}{3}$} u'' + 2 u^2) \, a^{-2 \gamma} + O(a^{-3 \gamma}) \label{4.19} \\
V_{1} & = & 4 w a^{-\gamma} + O(a^{-2 \gamma}) \label{4.20} \\
V_{2} & = & 8 u + O(a^{- \gamma}) \label{4.21} \\
V_{3} & = & O(1) \label{4.22}
\end{eqnarray}
This implies
\begin{eqnarray}
Q &=& \partial^3 + \frac{3}{2} \{u,\partial\} + 3w \label{4.23} \\
P &=& \partial^4 + 2 \{u,\partial^2 \} + 2 \{w , \partial \} - \frac{1}{3} u'' + 2 u^2 \label{4.24} 
\end{eqnarray}
and
\begin{eqnarray}
D_{1}(w,u) & = & w'' + 6 w u \nonumber \\
           & = &  \int r_{1} dx = h_{1} = 0\label{4.25}
\end{eqnarray}
These results are equivalent with those known \cite{7}.

The case $(l_{1},l_{2}) = (5,4)$ corresponds to the tricritical Ising model. We have two functions $w_{1}$, $w_{2}$ which we define by (\ref{4.15}) and
\begin{equation}
u_{-1} = -5 u_{1} + v_{-1} + 2 w_{1}\, a^{- \gamma} + (6 u_{1}^2 + v_{-1}^2 + w_{2}) a^{-2 \gamma}
\label{4.26}
\end{equation}
it follows then from (\ref{3.23}),(\ref{3.24})
\begin{eqnarray}
U_{0} & = & (u'' + 6 u^2 + 4 w_{2})\, a^{-2 \gamma} + O(a^{-3 \gamma})\label{4.27} \\
U_{1} & = & 4 w_{1}\, a^{- \gamma} + O(a^{-2 \gamma}) \label{4.28} \\
U_{2} & = & 8 u  + O(a^{- \gamma}) \label{4.29} \\
U_{3} & = & O(1) \label{4.30} \\
V_{0} & = & (\mbox{$- \frac{5}{4}$} w_{1}'' - 5 u w_{1})\, a^{-3 \gamma} + O(a^{-4 \gamma}) \label{4.31} \\
V_{1} & = & (\mbox{$- \frac{5}{2}$} u'' - 10 u^2 - 5 w_{2})\, a^{-2 \gamma} + O(a^{-3 \gamma}) \label{4.32} \\
V_{2} & = & - 10 w_{1}\, a^{- \gamma} + O(a^{-2 \gamma}) \label{4.33} \\
V_{3} & = & - 30 u  + O(a^{- \gamma}) \label{4.34} \\
V_{4} & = & O(1) \label{4.35} 
\end{eqnarray}
The two differential operators come out as
\begin{equation}
Q = \partial^4 + \{2 u,\partial^2 \} + \{2 w_{1}, \partial \} + 4 w_{2} + 6 u^2
\label{4.36}
\end{equation}
\begin{equation}
- P = \partial^5 +\frac{5}{2} \{u ,\partial^3 \} + \frac{5}{2} \{w_{1},\partial^2 \} + \frac{5}{2} \{w_{2} - \mbox{$\frac{1}{4}$} u'' + 2 u^2, \partial \} + 5 u w_{1}
\label{4.37}
\end{equation}
After the substitutions
\begin{equation}
u \rightarrow - \frac{1}{2} u
\label{4.38}
\end{equation}
\begin{equation}
w_{1} \rightarrow  \frac{1}{2} w
\label{4.39}
\end{equation}
\begin{equation}
4 w_{2} + 2 u^2 \rightarrow  v
\label{4.40}
\end{equation}
these differential operators and the differential equations (\ref{3.41}),
(\ref{3.44}) can be shown to be identical with those of Ginsparg et al. 
\cite{7}.

Now we come to the case $(l_{1},l_{2}) = (6,4)$ which was not studied before. In this case there is a common divisor of $l_{1}$ and $l_{2}$ : 2. Once again we can define $w_{1}$ and $w_{2}$ as in (\ref{4.15}),(\ref{4.26}) but we prefer to define $w_{2}$ b
y 
\begin{equation}
u_{-1} = -5 u_{1} + v_{-1} + 2 w_{1}\, a^{- \gamma} + (\mbox{$\frac{9}{2}$} u_{1}^2 - \mbox{$\frac{1}{2}$} v_{-1}^2 - 3 u_{1} v_{-1} + w_{2}) a^{-2 \gamma}
\label{4.41}
\end{equation}
because this minimizes the terms in the differential equations that we have to give explicitly. The differential operators are
\begin{equation}
Q = (\partial^2 + 2 u)^2 + 2 \{ w_{1}, \partial \} - 4 w_{2} 
\label{4.42}
\end{equation}
\begin{equation}
P = (\partial^2 + 2 u)^3 + 3 \{w_{1} ,\partial^3 \} - 3 \{w_{2},\partial^2 \} + \{ -\mbox{$\frac{1}{2}$} w_{1}'' + 6 u w_{1} , \partial \} - 2 w_{2}'' + 6 w_{1}^2- 12 u w_{2}
\label{4.43}
\end{equation}
By explicit calculation we found that
\begin{equation}
P = Q^{\frac{3}{2}}_{+}
\label{4.44}
\end{equation}
Moreover we get the differential equations
\begin{eqnarray}
h_{2} & = & D_{2}(w_{1},w_{2},u) \nonumber \\
      & = & - w_{1}^{(4)} - 8 u w_{1}'' - 12 u' w_{1}' - 4 u'' w_{1} - 24 w_{1} w_{2} \label{4.45} \\
h_{1} & = & D_{1}(w_{1},w_{2},u) \nonumber \\
      & = & w_{2}^{(4)} + 8 u w_{2}'' + 4 u' w_{2}' + 12 w_{2}^2 -10 w_{1} w_{1}''- 5 w_{1}'\,^2 - 24 u w_{1}^2 \label{4.46}\\
x & = & \int [ 2 r_{0} + \mbox{$\frac{1}{2}$} r_{2}'' + 2 u' ( D_{2}(w_{1},w_{2},u) - h_{2}) ] dx \nonumber \\
  & = & 8 w_{1} w_{2}'' - 8 w_{1}' w_{2}' - 16 w_{1}^3 - 2 h_{2} u \label{4.47}
\end{eqnarray}
To make $2 r_{0}$ integrable we had to add the term $2 u' ( D_{2} - h_{2})$. But this is also necessary in the (5,4) example (see \cite{7}).

\section{Examples from the series (B)}

The simplest non trivial example is the symmetric case 
\begin{displaymath}
(l_{1},l_{2}) = (3,3)
\end{displaymath}
There exists one function $w_{1}=w$ which we introduce by (\ref{4.15}). In this example we obtain from perturbation theory
\begin{equation}
- \, P^{(0)} = Q^{(0)} = \partial^3 + \frac{3}{2} \{u,\partial\} + 3w
\label{5.1}
\end{equation}
\begin{equation}
-\,P^{(1)} - Q^{(1)} = \partial^4 + 2 \{u,\partial^2 \} + 2 \{w , \partial \} - \frac{1}{3} u'' + 2 u^2
\label{5.2}
\end{equation}
and
\begin{equation}
D_{1}(w,u)   =  w'' + 6 w u = 0
\label{5.3}
\end{equation}
$P^{(1)}$ and $Q^{(1)}$ are separately not gauge invariant.

Thus we have reproduced the differential operators and equations (\ref{4.23})-(\ref{4.25}). The commutator
\begin{equation}
\left[\; \sum_{n} a^{-n \gamma} P^{(n)} , \sum_{m} a^{-m \gamma} Q^{(m)}\;\right] = \left[\,- P^{(1)} - Q^{(1)} , Q^{(0)}\,\right] a^{- \gamma} + O(a^{- 2 \gamma})
\label{5.4}
\end{equation}
gives an additionnal factor $a^{- \gamma}$ that  enlarges the total power of $a^{- \gamma}$ in $[B_{2},B_{1}]$ to
\begin{equation}
3 + 3 + 1 = 7 = 4 + 3
\label{5.5}
\end{equation}
as is necessary for the universality class $[4,3]$.

Now we turn to the rather complicated case 
 \begin{displaymath}
(l_{1},l_{2}) = (6,3)
\end{displaymath}
Again we introduce $w_{1}=w$ by (\ref{4.15}). The differential operators are  
\begin{eqnarray}
Q^{(0)} & = & \partial^3 + \frac{3}{2} \{u,\partial\} + 3w \label{5.6} \\
Q^{(1)} & = & - \frac{1}{2} \partial^4 - \frac {1}{4} \{u_{1}+ 7 v_{-1},\partial^2 \} - \{w , \partial \} + \frac{1}{24}(u_{1}'' + 13 v_{-1}'')  + 3 u_{1}^2 - u^2 \label{5.7} \\
P^{(0)} & = & (Q^{(0)})^2 \label{5.8}\\
P^{(1)} & = & 2 \partial^7 + \frac{1}{2} \{17 u_{1} + 11 v_{-1}, \partial^5 \} +  7 \{w , \partial^4 \} \nonumber \\
        &   & + \frac{1}{2} \{ - \mbox{$\frac{1}{12}$}(281 u_{1}'' + 161 v_{-1}'') + 43 u_{1}^2 + 56  u_{1} v_{-1} + 19 v_{-1}^2 , \partial^2 \} \nonumber \\
        &   & + \{ \mbox{$\frac{1}{12}$}(83 u_{1}'''' + 47 v_{-1}'''') - \mbox{$\frac{1}{8}$} (155 u_{1} u_{1}'' + 119 u_{1}''v_{-1} + 107 u_{1} v_{-1}'' + 71 v_{-1} v_{-1}'' ) \nonumber \\
        &   & - \mbox{$\frac{9}{4}$}(15 (u_{1}')^2  + 22 u_{1}' v_{-1}' + 7 (v_{-1}')^2) \nonumber \\
        &   & + 26 u_{1}^3 + 42 u_{1}^2 v_{-1} + 24 u_{1} v_{-1}^2 + 8 v_{-1}^3 + 30 w^2 \;,\; \partial \} \nonumber \\
        &   & + \frac{11}{6} w'''' + 3 u w'' - \frac{1}{2} w' ( 49 u_{1}' + 31 v              _{-1}') -  \frac{1}{4} w ( 15 u_{1}'' + 3 v_{-1}'')\nonumber \\
        &   & + 18 w (3 u_{1}^2 + 4 u_{1} v_{-1} + 2 v_{-1}^2) \label{5.9} 
\end{eqnarray}
Both $P^{(1)}$ and $Q^{(1)}$ are gauge dependent. We consider the commutator
\begin{eqnarray}
& & \left[\; \sum_{n} a^{-n \gamma} P^{(n)}\; , \;\sum_{m} a^{-m \gamma} Q^{(m)}\;\right] = \nonumber\\ 
  & & \left[\; \sum_{n} a^{-n \gamma} P^{(n)} - (\;\sum_{m}a^{-m \gamma} Q^{(m)})^2 \;, \; \sum_{m} a^{-m \gamma} Q^{(m)}\;\right] = \nonumber \\
  & & \left[\; P^{(1)} - Q^{(0)}Q^{(1)} - Q^{(1)}Q^{(0)} , Q^{(0)} \;\right]a^{- \gamma} + O(a^{-2 \gamma}) \label{5.10} 
\end{eqnarray}
The first differential operator in (\ref{5.10}) can be calculated and is
\begin{equation}
P^{(1)} - (Q^{(0)}Q^{(1)} + Q^{(1)}Q^{(0)}) = 3 (Q^{(0)})^{\frac{7}{3}}_{+}
\label{5.11}
\end{equation}
The factor 3 can be eliminated by rescaling as was discussed in section 3. The additional factor $a^{- \gamma}$ changes the total power of $a^{- \gamma}$ in $[B_{2}, B_{1}]$ to
\begin{equation}
6 + 3 + 1 = 7 + 3
\label{5.12}
\end{equation}
as is necessary for the universality class $[7,3]$. Thus this class has been proven to arise.

We remember that one-matrix models lead to universality classes $[2m+1 , 2]$, and we could imagine that such classes can also be produced from two-matrix models with singularities
\begin{displaymath}
(l_{1},l_{2}) = (2m , 2)  
\end{displaymath}
However , if we insert an ansatz for such singularity into the Dyson-Schwinger equations, the perturbative solution cannot be made to satisfy the commutator equation
\begin{equation}
[B_{2}, B_{1}] = 1
\label{5.13}
\end{equation}
at any order in $a^{- \gamma}$ as we shall show in the sequel.

Of course a quadratic potential $V_{2}$ can be eliminated from the partition function by performing a Gaussian integral. In the Dyson-Schwinger equations this comes out as follows. If
\begin{eqnarray}
A_{1} + c B_{2} & = & V_{1}'(B_{1})  \label{5.14} \\
A_{2} + c B_{1} & = & g_{1}^{(2)} + g_{2}^{(2)} B_{2} \;, \;\; (g_{2}^{(2)}\neq 0) \label{5.15}
\end{eqnarray}
then $B_{2}$ can be eliminated by substituting (\ref{5.15}) into (\ref{5.14})
\begin{eqnarray}
A_{1} + \frac{c}{g_{2}^{(2)}} A_{2} & = & \frac{c}{g_{2}^{(2)}}(g_{1}^{(2)}- c B_{1}) + V_{1}'(B_{1}) \nonumber \\
 & = & \tilde{V}(B_{1}) \frac{}{} \label{5.16}
\end{eqnarray}
Now $A_{1}$ is strictly lower and $A_{2}$ strictly upper triangular and under transposition
\begin{equation}
B_{1} \rightarrow B_{1}^{T}
\label{5.17}
\end{equation}
they exchange their role in (\ref{5.16}). In the one-matrix model there is only one orthogonal system $\{\Pi_{m}(\lambda)\}$ (section 2) implying that $B_{1}$ is symmetric. In this case (\ref{5.16}) reduces to
\begin{equation}
A_{1} + A_{1}^{T} = \tilde{V}'(B_{1}) 
\label{5.18}
\end{equation}
Via the antisymmetric matrix
\begin{equation}
C = A_{1} - \frac{1}{2} \tilde{V}'(B_{1})
\label{5.19}
\end{equation}
and the double scaling limit
\begin{eqnarray}
B_{1} & \rightarrow & a^{-2 \gamma} Q \label{5.20} \\
C & \rightarrow & a^{-(2m+1) \gamma} P \label{5.21}
\end{eqnarray}
the universality class $[2m + 1 , 2]$ is obtained.

Instead we solve the Dyson-Schwinger equations (\ref{5.14}), (\ref{5.15}) perturbatively as they stand. First we set 
\begin{equation}
L_{1} = l_{1} = 4 \;, \; m = 2
\label{5.22}
\end{equation}
Our algorithm simplifies by the equalities to all orders
\begin{eqnarray}
u_{0}(x) & = & v_{0}(x) \label{5.23} \\
u_{-1}(x)& = & v_{-1}(x) \label{5.24}
\end{eqnarray}
Due to the fact that no functions $\{w_{k}\}$ are needed, we obtain only even powers in the expansions (\ref{1.1}), (\ref{1.2})
\begin{equation}
B_{1}  \rightarrow  a^{-2 \gamma}\sum_{n=0}^{\infty} a^{-2n\gamma} Q^{(2n)} 
\label{5.25} 
\end{equation}
\begin{equation}
B_{2}  \rightarrow  a^{-4 \gamma}\sum_{n=0}^{\infty} a^{-2n\gamma} P^{(2n)} 
\label{5.26}
\end{equation}
We find 
\begin{eqnarray}
Q^{(0)} & = & \partial^2 + 2 u \label{5.27} \\
Q^{(2)} & = &  \frac{1}{12} \partial^4 + \frac {1}{4} \{u ,\partial^2 \} - \frac{1}{8} u'' - \frac{1}{4} (u_{1} - v_{-1})^2 \label{5.28} \\
P^{(0)} & = & (Q^{(0)})^2 \label{5.29}\\
P^{(2)} & = & \frac{2}{3} \partial^6 + \frac{1}{12} \{41 u_{1} + 11 v_{-1}, \partial^4 \}  \nonumber \\
        &   & + \{ - \mbox{$\frac{1}{8}$}(23 u_{1}'' + 3 v_{-1}'') + 16 u_{1}^2 + 15  u_{1} v_{-1} +  v_{-1}^2 , \partial^2 \} \nonumber \\
        &   & +  \frac{1}{4}(5 u_{1}''' +  v_{-1}''') - \frac{1}{2}(15 u_{1}'^{2} + 9 u_{1}'v_{-1}') \nonumber\\ 
        &   & + \frac{1}{8} ( -23 u_{1}u_{1}'' + 5 u_{1}v_{-1}'' - 5 v_{-1}u_{1}'' + 7 v_{-1}v_{-1}'') \nonumber \\
        &   & + \frac{1}{4}(27 u_{1}^3 + 67 u_{1}^{2}v_{-1} + 37 u_{1} v_{-1}^{2} - 3 v_{-1}^{3}) \label{5.30}
\end{eqnarray}

An almost trivial observation is that all $P^{(2n)}$, $Q^{(2n)}$ are symmetric operators. Consequently the commutator
\begin{eqnarray}
& & \left[\; \sum_{n} a^{-2n \gamma} P^{(2n)}\; , \;\sum_{m} a^{-2m \gamma} Q^{(2m)}\;\right] = \nonumber\\ 
  & & \left[\; P^{(2)} - Q^{(0)}Q^{(2)} - Q^{(2)}Q^{(0)} , Q^{(0)} \;\right]a^{-2 \gamma} + O(a^{-4 \gamma})=  \nonumber \\
  & & \sum_{k=0}^{5} \{ r_{k}^{(2)} , \partial^{k} \}\,a^{-2 \gamma} + O(a^{-4 \gamma}) \label{5.31}   
\end{eqnarray}
is antisymmetric
\begin{equation}
r_{0}^{(2)} = r_{2}^{(2)} = r_{4}^{(2)} = 0 
\label{5.32}
\end{equation}
This symmetry argument is valid to any order (and for any m)
\begin{equation}
r_{0}^{(2n)} = 0 \;,\; \mbox{all n}
\label{5.33}
\end{equation}
and therefore the commutator relation cannot be satisfied.

Only as a side remark we add that the three differential equations
\begin{equation}
r_{1}^{(2)} = r_{3}^{(2)} = r_{5}^{(2)} = 0 
\label{5.34}
\end{equation}
have the unique solution
\begin{equation}
u_{1} = v_{-1}
\label{5.35}
\end{equation}
which implies
\begin{equation}
B_{1} = B_{1}^{T}
\label{5.36}
\end{equation}

\section{Expectation values and gauge invariance}

The basis functions $\{ w_k \}$ have hitherto been introduced (see (\ref{4.15}),
(\ref{4.26}),(\ref{4.41})) by the requirement that the relevant objects,
namely differential operators and equations are gauge invariant. This fixes
them only at low orders. Expectation values are a simple device to extend this
definition to higher orders.

For any two-matrix model we obtain for a finite $N$ using (\ref{2.11}),
(\ref{2.17}),(\ref{2.28}) and (\ref{3.26}), (\ref{3.27}) that the partition
function can be expressed as
\begin{equation} \label{6.1}
Z = C(N) N! (s_0 {\tilde{s}}_0)^{-N} \prod_{k=1}^{N-1}(r_{1}(N-k-1)
s_{-1}(N-k))^{k}
\end{equation}
so that the free energy is
\begin{eqnarray}
F &=& \log Z = \log(C(N) N!)- N \log(s_0 {\tilde{s}}_0) \nonumber \\
&+& a^{-2 \gamma} (1-a^2 z)^{-2}\int_{a^{-2}}^{z} dx (z-x) 
\log(1+a^{-2 \gamma} u(x)) \nonumber \\
\label{6.2} &+& \mbox{ corrections from Euler's summation formula.}
\end{eqnarray}
The latter corrections in (\ref{6.2}) form an infinite series. The function
$u(x)$ increases for $x \rightarrow \infty$ and one can convince oneself
that
\begin{equation} \label{6.3}
a^{-2 \gamma} u(a^{-2}) = O(1)
\end{equation}
so that the perturbative expansion of the integral in (\ref{6.2}) fails at
the lower end. The corrections due to Euler's summation formula contain 
terms
\begin{equation} \label{6.4}
u^{(n)}(a^{-2}), \quad n \geq 0.
\end{equation}
Therefore we set
\begin{equation} \label{6.5}
F(z) = F(0) + (F(z)-F(0))
\end{equation}
and expand only the difference perturbatively. A similar caveat must be applied
to the expectation values.

Expectation values
\begin{equation} \label{6.6}
\langle G(M^{(1)}, M^{(2)}) \rangle
\end{equation}
with respect to the partition function (\ref{2.5}) or (\ref{2.11}) can 
apparently be obtained by parametric differentiation, if $G$ is a polynomial
of the variables
\begin{equation} \label{6.7}
\mbox{Tr}(M^{(1)})^{a}, \quad \mbox{Tr}(M^{(2)})^{b}, \quad 
\mbox{Tr}(M^{(1)}M^{(2)}).
\end{equation}
After reduction of the partition function by means of Mehta's theorem, these
variables correspond to
\begin{equation} \label{6.8}
\sum_{i} \lambda_{i}^{a}, \quad \sum_{i} \mu_{i}^{b}, \quad \sum_{i} \lambda_{i}
\mu_{i}, \quad (\lambda_{i}=\lambda_{i}^{(1)}, \mu_{i}=\lambda_{i}^{(2)}).
\end{equation}
First we show that the expectation value of the third variable with a reduced
partition function is not gauge invariant.

Let us denote the trace over the upper left $N \times N$ submatrix of an
infinite ($\bbbn \times \bbbn$) matrix by ${\mbox{Tr}}_{\scriptsize{N}}$. 
Then insertion
of
\[ \sum_{i} \lambda_{i} \mu_{i} \]
into (\ref{2.11}) gives for fixed $N$
\[ {\mbox{Tr}}_{\scriptsize{N}}(B_1 B_2) \]
which can be evaluated to yield
\begin{eqnarray}
&N& {l_1+l_2 \choose l_1} - \left[ \frac{(l_1+l_2-1)!}{l_1! l_2!} (l_1 l_2
-l_1 -l_2) +1 \right] \nonumber \\
\label{6.9} &-& a^{-\gamma} \int_{a^{-2}}^{z} dx \left\{ \sum_{k}
[\rho_{-k} v_{k}(x)+ \sigma_{k} u_{-k}(x)] \right\} +\mbox{O}(a^{-3 \gamma})
\end{eqnarray}
for any maximal singularity $(l_1, l_2)$. For $(l_1, l_2)=(4,3)$ the integral
is to leading order (using still (\ref{4.15}))
\begin{equation} \label{6.10}
14 u_{1}(x) - 43 v_{-1}(x) - 17 w_{1}(x) a^{-\gamma}
\end{equation}
which is not gauge invariant. Therefore only the first two expressions in
(\ref{6.8}) make sense.

Expectation values of monomials can be reduced easily (see \cite{19})
\begin{eqnarray}
\label{6.11} \langle \mbox{Tr}(M^{(1)})^{a} \mbox{Tr}(M^{(1)})^{b} \rangle &=&
{\mbox{Tr}}_{\scriptsize{N}}(B_{1}^{a}) {\mbox{Tr}}_{\scriptsize{N}}(B_{1}^{b})
+{\mbox{Tr}}_{\scriptsize{N}}(B_{1}^{a} \Pi_N B_{1}^{b}) \\
\label{6.12} \langle \mbox{Tr}(M^{(1)})^{a} \mbox{Tr}(M^{(2)})^{b} \rangle &=&
{\mbox{Tr}}_{\scriptsize{N}}(B_{1}^{a}) {\mbox{Tr}}_{\scriptsize{N}}(B_{2}^{b})
+{\mbox{Tr}}_{\scriptsize{N}}(B_{1}^{a} \Pi_N B_{2}^{b})
\end{eqnarray}
where $\Pi_{N}$ is the projection matrix
\begin{equation} \label{6.13}
(\Pi_N)_{rs} = \left\{ \begin{array}{l} \delta_{rs} \; \mbox{ if } r \geq N 
\mbox{ and } s \geq N \\ 0 \quad \mbox{ else}. \end{array} \right. 
\end{equation}
Of course for the simple cases of one variable (\ref{6.7}) we have from
(\ref{6.11}), (\ref{6.12})
\begin{eqnarray}
\label{6.14} \langle \mbox{Tr}(M^{(1)})^{a} \rangle &=& 
{\mbox{Tr}}_{\scriptsize{N}}(B_{1})^{a} \\
\label{6.15} \langle \mbox{Tr}(M^{(2)})^{b} \rangle &=& 
{\mbox{Tr}}_{\scriptsize{N}}(B_{2})^{b}.
\end{eqnarray}
Consider (\ref{6.14}) with $a=1$. Then
\begin{eqnarray}
{\mbox{Tr}}_{\scriptsize{N}} B_{1} &=& N \rho_{0} - a^{-\gamma} 
\int_{a^{-2}}^{z} dx u_{0}(x) \nonumber \\
\label{6.16} &+& \mbox{ corrections from Euler's summation formula}.
\end{eqnarray}
We recognize that instead of (\ref{4.15}) we better use
\begin{eqnarray}
u_{0}(x) &=& u_1 + v_{-1} + a^{-\gamma} w_{1} + a^{-2 \gamma} u_1 v_{-1}
\nonumber \\
\label{6.17} &=& u + a^{-\gamma} w_1 .
\end{eqnarray}
The same argument can be applied to $M^{(2)}$
\begin{eqnarray}
{\mbox{Tr}}_{\scriptsize{N}} B_{2} &=&  N \sigma_{0} - a^{-\gamma} 
\int_{a^{-2}}^{z} dx v_{0}(x) \nonumber \\
\label{6.18} &+& \mbox{ corrections from Euler's summation formula}.
\end{eqnarray}
so that $v_{0}$ must depend only on $u$ and $w_1$. Indeed we find for the
maximal singularity $(4,3)$ from (\ref{6.17})
\begin{equation} \label{6.19}
v_{0} = u -w_1 a^{-\gamma}-(\mbox{$\frac{1}{12}$} u'' + u^2) a^{-2 \gamma} +
\mbox{$\frac{2}{5}$} (\mbox{$\frac{1}{6}$} w_1'' +u w_1) a^{-3 \gamma} + 
\mbox{O}(a^{-4 \gamma})
\end{equation}
and for $(3,3)$
\begin{equation} \label{6.20}
v_{0} = u -w_1 a^{-\gamma}-(\mbox{$\frac{1}{12}$} u'' + u^2) a^{-2 \gamma} +
2 (\mbox{$\frac{1}{6}$} w_1'' +u w_1) a^{-3 \gamma} + 
\mbox{O}(a^{-4 \gamma}).
\end{equation}
Finally we consider the cases $a=2$ in (\ref{6.14}) and $b=2$ in (\ref{6.15}).

To sum up the results we introduce abbreviations
\begin{eqnarray}
\label{6.21} \phi_{1}(x) &=& u(x) \\
\label{6.22} \phi_{2}(x) &=& w_{1}(x) \\
\label{6.23} \phi_{3}(x) &=& \mbox{$\frac{1}{12}$} u''(x) + u^{2}(x) \\
\label{6.24} \phi_{3}(x) &=& \mbox{$\frac{1}{6}$} w_1''(x) +u(x) w_{1}(x)
\end{eqnarray}
and
\begin{equation} \label{6.25}
\Phi_{n}(z) = \int_{a^{-2}}^{z} \phi_{n}(x) \, dx .
\end{equation}
We write for (\ref{6.14}), (\ref{6.15})
\begin{equation} \label{6.26}
\gamma_1 N + \gamma_0 + \sum_{n \geq 1} c_n \Phi_{n}(z) a^{-n \gamma}
\end{equation}
and obtain the coefficients in the following table \\[0.8cm]
\hspace*{3.8cm} \begin{tabular}{|c|cc|cccccc|} \hline
sing & a & b & $\gamma_1$ & $\gamma_0$ & $c_1$ & $c_2$ & $c_3$ & $c_4$ \\ \hline 
 (4,3) & 1 & & -3 & 0 & -1 & -1 & 0 & 0 \\
 & 2 & & 15 & -3 & 4 & 2 & -1 & $-\frac{22}{5}$ \\
 & & 1 & -4 & 0 & -1 & 1 & 1 & $-\frac{2}{5}$ \\
 & & 2 & 28 & -6 & 6 & -4 & -5 & -10 \\ \hline
 (3,3) & 1 & & -3 & 0 & -1 & -1 & 0 & 0 \\
 & 2 & & 15 & -3 & 4 & 2 & -1 & 2 \\
 & & 1 & -3 & 0 & -1 & 1 & 1 & -2 \\
 & & 2 & 15 & -3 & 4 & -2 & -3 & 18 \\ \hline
 \end{tabular} \\[0.35cm]
\hspace*{3.8cm} \parbox{7.6cm}{Table 1. Expansion coefficients for the expectation 
values of (\ref{6.14}), (\ref{6.15}) for the singularities $(4,3)$ and $(3,3)$.}
\\[0.5cm]
\noindent The general expression for $\gamma_1$ is
\renewcommand{\arraystretch}{2}
\begin{equation} \label{6.27}
\gamma_1 = \left\{ \begin{array}{l} (-1)^{a} {a \, l_2 \choose a}, \mbox{ respectively} \\ 
(-1)^{b} \, {b \: l_1 \choose b}. \end{array} \right.
\end{equation}

\end{document}